\documentclass[10pt]{article}
\usepackage{graphicx}

\def\Title#1{\begin{center} {\Large #1 } \end{center}}
\def\Author#1{\begin{center}{ \sc #1} \end{center}}
\def\Address#1{\begin{center}{ \it #1} \end{center}}

\newcommand\pubblock{\rightline{\begin{tabular}{l} Proceedings of the Fifth Annual LHCP\\ \pubnumber\\
         \pubdate  \end{tabular}}}

\newenvironment{Abstract}{\begin{quotation} \begin{center} 
             \large ABSTRACT \end{center}\bigskip 
      \begin{center}\begin{large}}{\end{large}\end{center} \end{quotation}}

\newenvironment{Presented}{\begin{quotation} \begin{center} 
             PRESENTED AT\end{center}\bigskip 
      \begin{center}\begin{large}}{\end{large}\end{center} \end{quotation}}

\def\Acknowledgements{\bigskip  \bigskip \begin{center} \begin{large}
             \bf ACKNOWLEDGEMENTS \end{large}\end{center}}

\def\beq{\begin{equation}}
\def\eeq#1{\label{#1}\end{equation}}
\def\eeqn{\end{equation}}

\def\beqa{\begin{eqnarray}}
\def\eeqa#1{\label{#1}\end{eqnarray}}
\def\eeqan{\end{eqnarray}}

\let\bar=\overbar

\def\Dslash{\not{\hbox{\kern-4pt $D$}}}
\def\dslash{\not{\hbox{\kern-2pt $\del$}}}

\def\msb{{\bar{\ssstyle M \kern -1pt S}}}

\usepackage{color}
\usepackage{amsmath}

\newcommand\pubnumber{ CMS CR-2017/306 }

\newcommand\pubdate{\today}

\def\affiliation{
On behalf of the CMS collaboration \\
State Key Laboratory of nuclear Physics and Technology, Peking University \\
100871 Beijing, China}

\begin{document}

\large
\begin{titlepage}
\pubblock

\vfill
\Title{  Searches for diboson resonances at CMS  }
\vfill

\Author{ HUANG HUANG  }
\Address{\affiliation}
\vfill
\begin{Abstract}

A list of searches is presented for massive resonances decaying to WW, WZ, ZZ, WH, ZH and HH boson pairs in proton-proton collision data collected by the CMS experiment at the LHC.
The data are taken at centre-of-mass energies of $13~\mathrm{TeV}$, corresponding to respective integrated luminosities of $12.9~\mathrm{fb}^{-1}$(ICHEP) and up to $35.9~\mathrm{fb}^{-1}$.
The results are interpreted in the context of heavy vector triplet and singlet models that mimic properties of composite-Higgs models predicting $W'$ and $Z'$ bosons decaying to WZ, WW, WH, and ZH bosons.
A model with a bulk graviton that decays into WW and ZZ is also considered. Upper limits are set on the production cross section of the two models and no evidence is found for a signal.

\end{Abstract}
\vfill

\begin{Presented}
The Fifth Annual Conference\\
 on Large Hadron Collider Physics \\
Shanghai Jiao Tong University, Shanghai, China\\ 
May 15-20, 2017
\end{Presented}
\vfill
\end{titlepage}
\def\thefootnote{\fnsymbol{footnote}}
\setcounter{footnote}{0}
%

\normalsize 


\section{Introduction}

There are several theoretical models that motivate the existence of heavy particles that decay to
pairs of bosons. These models usually aim to answer open questions of the Standard Model (SM) such as the integration of gravity into the SM using extra dimensions.
Popular examples of such models include
the bulk scenario~\cite{Fitzpatrick:2007qr} of
the Randall-Sundrum Warped Extra Dimensions
model~\cite{Randall:1999ee}, and the composite heavy vector triplet (HVT) model~\cite{Pappadopulo:2014qza}.
The composite HVT generalizes a large number of explicit models predicting spin-1 resonances, which can be described by a rather small set of parameters.

In the bulk option, coupling of the graviton to light fermions is highly
suppressed, and the decay into photons is negligible,
while in the RS1 scenario, the graviton decays to photon and fermion pairs dominate.
In the context of WW and ZZ resonance searches, the bulk option is of great interest, since RS1 is already strongly constrained through searches in final states with fermions and photons.
The production of gravitons at hadron colliders in the bulk option is dominated by gluon-gluon fusion, and the branching fraction $\mathcal{B}(\rm G_{bulk} \to WW) \approx 2 \times \mathcal{B}(\rm G_{bulk} \to ZZ)$.
The decay mode into a pair of Higgs bosons has a branching fraction comparable to $\mathcal{B}(\rm G_{bulk} \to ZZ)$.

The $W'$ and $Z'$ bosons couple to the fermions through the combination of parameters $g^2 c_\mathrm{F}/ g_\mathrm{V}$ and to the H and vector bosons through $g_\mathrm{V} c_\mathrm{H}$ , where $g$ is the SU(2)$_L$ gauge coupling.

The production of $W'$ and $Z'$ bosons at hadron colliders is expected to be dominated by the process $qq' \to W' $ or $Z'$.
Two benchmark models are studied, denoted A and B. In model A, weakly coupled vector resonances arise from an extension of the SM gauge group. 
In model B, the heavy vector triplet is produced strongly coupled, such as in the composite Higgs model.
Consequently, in model A the branching fractions to fermions and SM massive bosons are comparable, whereas in model B, fermionic couplings are suppressed.
Therefore, in the context of WW, WZ, ZH, and WH resonance searches, model B is of more interest, since model A is strongly constrained by searches in final states with fermions.
In both options, the heavy resonances couple as SM custodial triplets, so that $W'$ and $Z'$ are expected to be degenerate in mass, and the branching fractions $\mathcal{B}(W' \to WH)$ and $\mathcal{B}(Z' \to ZH)$ to be comparable to $\mathcal{B}(W' \to WZ)$ and $\mathcal{B}(Z'\to WW)$. And then, the bosons $W$, $Z$ and $H$ would decay to leptons, neutrinos and quarks, which will combine to plethora of final states, each one with its own peculiarities. See Table \ref{tab:final states}.

\begin{table}[b]
\begin{center}
\begin{tabular}{l|ccccc}  
&$V\to q\bar{q}$&$W\to l\nu$&$Z\to ll$&$Z\to \nu\nu$&$H\to b\bar{b}$ \\ \hline
$V\to q\bar{q}$     &$VV\to q\bar{q}q\bar{q}$& -- & -- & --& --\\
$W\to l\nu$            &$WV\to  l\nu q\bar{q}$& & -- & -- & -- \\
$Z \to ll$                &$ZV\to llq\bar{q}$& & & -- & -- \\
$Z\to\nu\nu$          & & &$ZZ\to ll\nu\nu$& & -- \\
$H\to b\bar{b} $   &$VH\to q\bar{q}b\bar{b}$&$WH\to l\nu b\bar{b} $&$ZH\to llb\bar{b} $&$ZH\to \nu\nu b\bar{b} $&$HH\to b\bar{b} b\bar{b} $ \\ \hline
\end{tabular}
\caption{ Plethora of final states, each one with its own peculiarities. }
\label{tab:final states}
\end{center}
\end{table}

\section{CMS Detector}
The central feature of the CMS apparatus is a superconducting solenoid of 6m internal diameter, providing a magnetic field of 3.8T. Contained within the superconducting solenoid volume are a silicon pixel and strip tracker, a lead tungstate crystal electromagnetic calorimeter (ECAL), and a brass and scintillator hadron calorimeter (HCAL), each composed of a barrel and two endcap sections. Muons are measured in gas-ionization detectors embedded in the steel flux-return yoke outside the solenoid. Extensive forward calorimetry complements the coverage provided by the barrel and endcap detectors. A more detailed description of the CMS detector, together with a definition of the coordinate system used and the relevant kinematic variables, can be found in Ref.~\cite{Chatrchyan:2008aa}.

\section{Reconstruction and Identification technology}
\subsection{Grooming and jet mass}
The event reconstruction employs a particle-flow (PF) algorithm, which uses an optimized combination of information from the various elements of the CMS detector to reconstruct and identify individual particles produced in each collision. The algorithm identifies each reconstructed particle either as an electron, a muon, a photon, a charged hadron, or a neutral hadron.
The PF candidates are clustered into jets using the anti-$k_{T}$ algorithm with a distance parameter $R=0.8$, after passing the charged-hadron subtraction (CHS) pileup mitigation algorithm~\cite{CMS:2014ata}.
For each event, a primary vertex is identified as the one with the highest sum of the $p_{T}^2$ of the associated reconstructed objects, jets and identified leptons, and missing transverse momentum. The CHS algorithm removes charged PF candidates with a track longitudinal impact parameter not compatible with this primary vertex.
The contribution to a jet of neutral particles originating from pileup interactions, assumed to be proportional to the jet area, is subtracted from the jet energy.
Jet energy corrections as a function of the $p_{T}$ and $\eta$ are extracted from simulation and data in dijet, multijet, $\gamma$+jets, and leptonic Z+jets events.
Jets are required to pass identification criteria in order to remove spurious jets arising from detector noise. This requirement has negligible impact on the signal efficiency.

Although AK8 CHS jets are considered for their kinematic properties, the mass of the jet and the substructure variables are determined with a more sophisticated algorithm than the CHS procedure, denoted as pileup-per-particle identification (PUPPI) \cite{Bertolini:2014bba}. The PUPPI algorithm uses a combination of the three-momenta of the particles, event pileup properties, and tracking information in order to compute a weight, assigned to charged and neutral candidates, describing the likelihood that each particle originates from a pileup interaction.
The weight is used to rescale the particle four-momenta, superseding the need for further jet-based corrections.
The PUPPI constituents are subsequently clustered with the same algorithm used for CHS jets, and then are matched with near 100\% efficiency to the AK8 jets clustered with the CHS constituents.

\begin{figure}[!hbtp]\centering
  \includegraphics[width=0.32\textwidth]{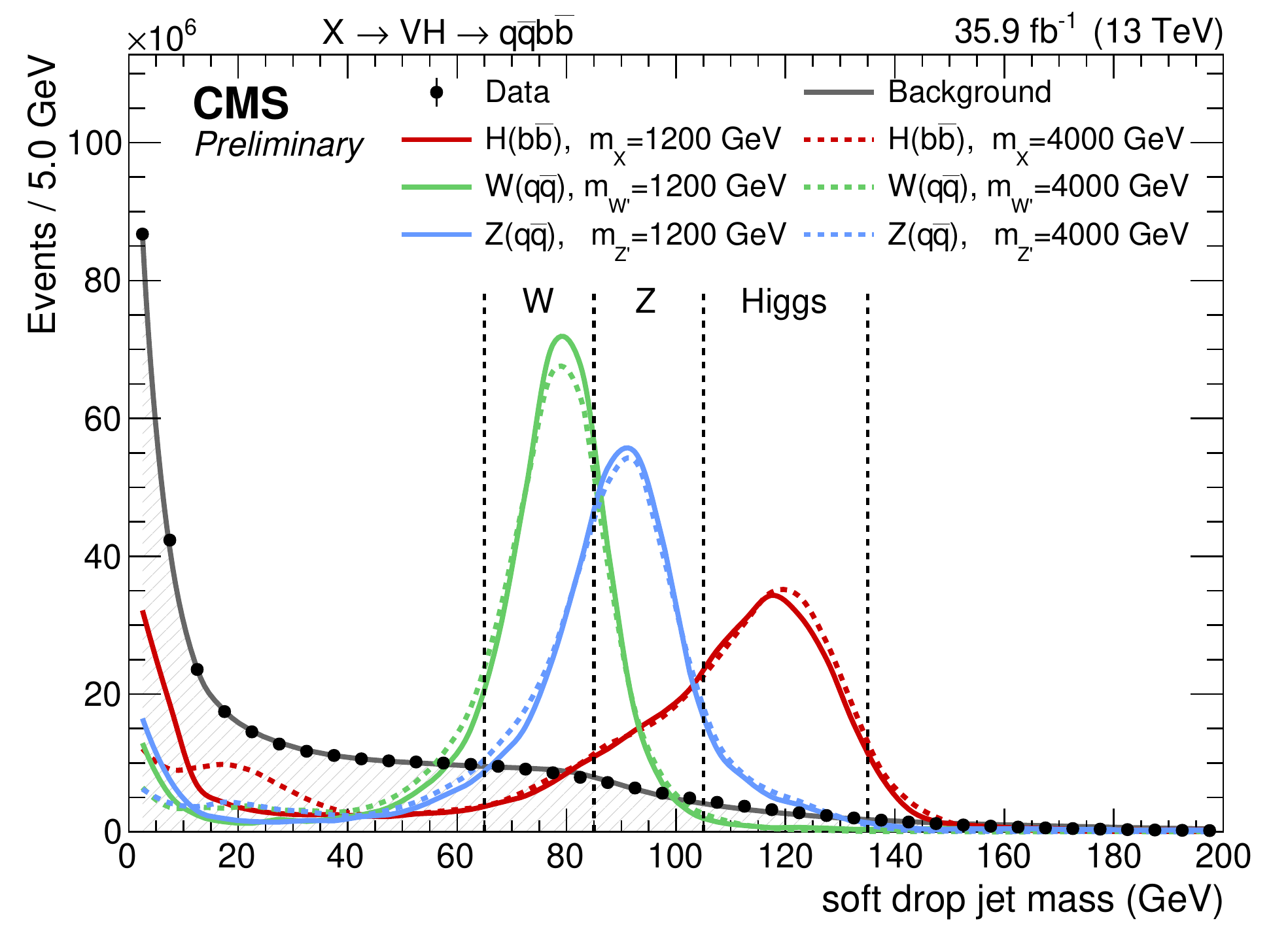}
  \includegraphics[width=0.32\textwidth]{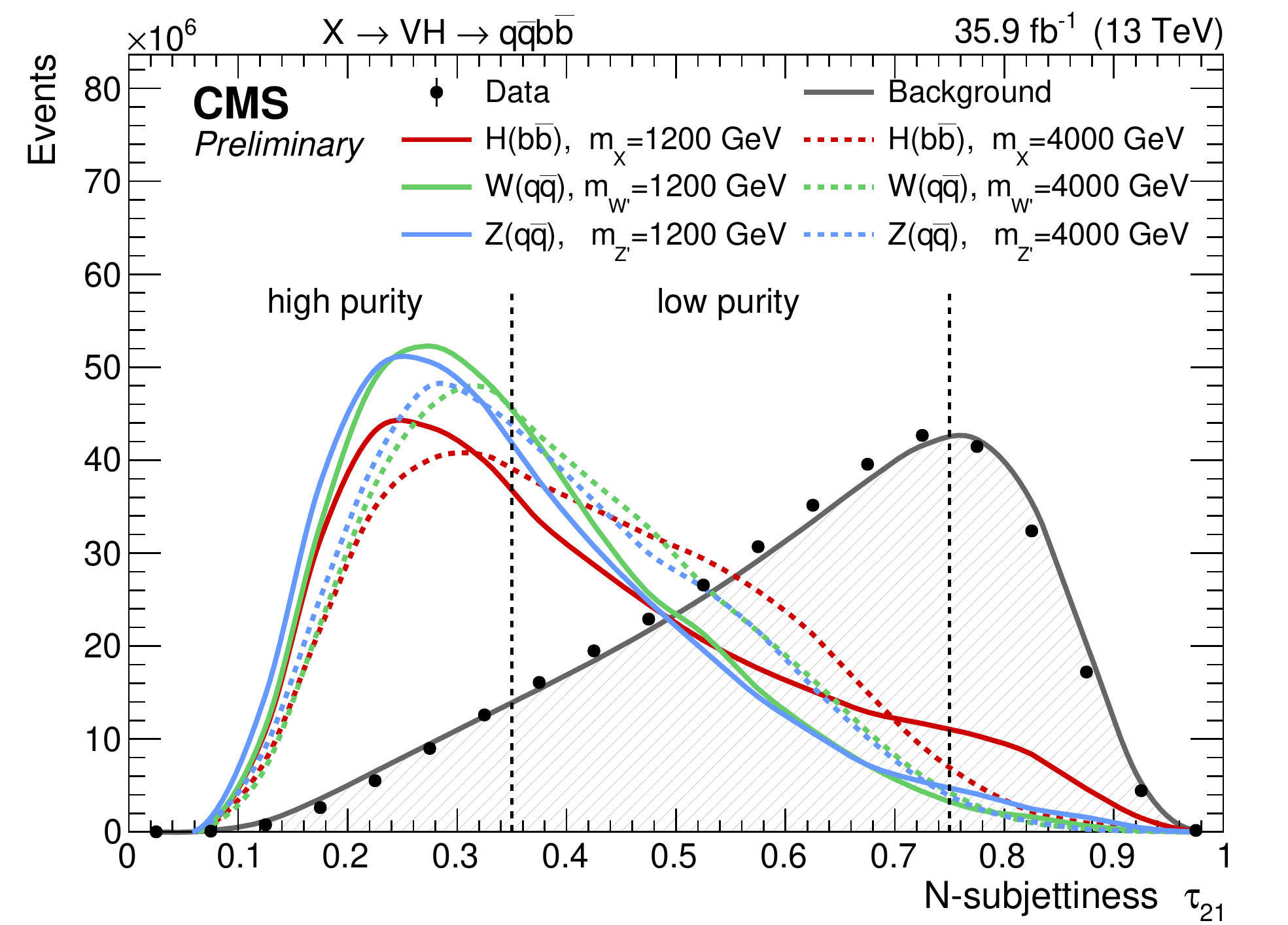}
  \includegraphics[width=0.32\textwidth]{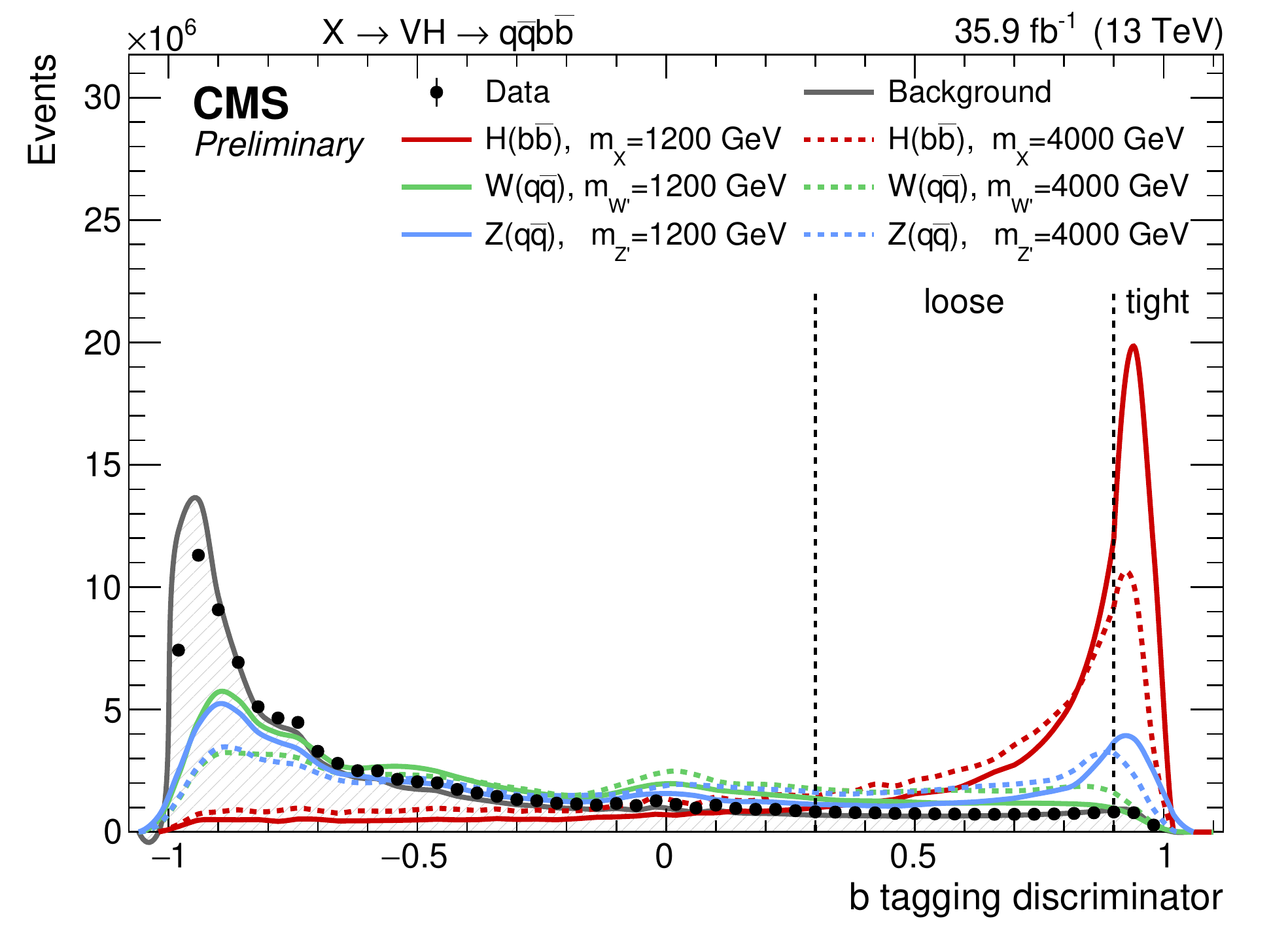}
  \caption{
    The left plot shows the distribution of the soft-drop PUPPI mass after the kinematic selections on the two jets, for data, simulated background, and signal. The signal events with low mass correspond to boson decays where one of the two quarks is emitted outside the jet cone or the two quarks are overlapping. The distributions are normalized to the number of events observed in data. The dashed vertical lines represent the boundaries between the jet mass categories. Distribution of the $N$-subjettiness $\tau_{21}$ (middle) and b tagging discriminator output (right) after the kinematic selections on the two jets, for data, simulated background, and signal. The distributions are normalized to the number of events observed in data. The dashed vertical lines represent the boundaries between the categories as described in the text.\cite{CMS:2017eme}
    \label{fig:tau21bb} }
\end{figure}

The soft-drop algorithm~\cite{Larkoski:2014wba}, which is designed to remove contributions from soft radiation and additional interactions, is applied to PUPPI jets. The angular exponent parameter of the algorithm is set to $\beta = 0$, and the soft threshold to $z_{cut} = 0.1$. The soft-drop jet mass is defined as the invariant mass associated with the four-momentum of the jet after the application of the soft-drop algorithm. See Figure.~\ref{fig:tau21bb} left plot.

Dedicated mass corrections, derived from simulation and data in a region enriched with $t\bar{t}$ events having merged $W(qq')$ decays, are applied to each jet mass in order to remove any residual $jet$ $p_{T}$ dependence, and to match the jet mass scale and resolution observed in data. The measured jet mass resolution, obtained after applying the PUPPI and soft-drop algorithms, is approximately 10\%.

\subsection{$N$-subjettiness and Vector boson tagging ($V\to q\bar{q}$)}
Substructure variables are used to identify single reconstructed jets that result from the merger of more than one parton jet. These variables are calculated on each reconstructed jet before the application of the soft-drop algorithm including the PUPPI algorithm corrections for pileup mitigation. The constituents of the jet are clustered iteratively with the anti-$k_{T}$ algorithm, and the procedure is stopped when $N$ subjets are obtained. A variable, the $N$-subjettiness~\cite{Thaler:2010tr}, is introduced:
$$\tau_N = \frac{1}{d_0} \sum_k p_{\mathrm{T},k} \, \min( \Delta R_{1,k}, \Delta R_{2,k}, \dots, \Delta R_{N,k} ).$$
The index $k$ runs over the jet constituents and the distances $\Delta R_{J,k}$ are calculated with respect to the axis of the $J$th subjet.
The normalization factor $d_0$ is calculated as $d_0 = \sum_k p_{\mathrm{T}, k} R_0$, setting $R_0$ to the radius of the original jet.
The variable that best discriminates between quark and gluon jets and jets from two-body decays of massive particles is the ratio of 2-subjettiness and 1-subjettiness, $\tau_{21} = \tau_{2} / \tau_{1}$, which lies in the interval from 0 to 1, where small values correspond to a high compatibility with the hypothesis of a massive object decaying into two quarks. See Figure.~\ref{fig:tau21bb} middle plot.
The normalization scale factors relative to the $\tau_{21}$ categories are measured from data in a sample enriched in $t\bar{t}$ events in two $\tau_{21}$ intervals 
These two selections are approximately 50 and 45\% efficient for identifying two-pronged jets produced in a decay of a massive boson, and 10 and 60\% efficient on one-pronged jets, respectively. The threshold values are chosen in order to maximize the overall sensitivity over the entire mass spectrum.

\subsection{Higgs boson tagging ($H\to b\bar{b}$)}
The Higgs boson jet candidates are identified using a dedicated b tagging discriminator, specifically designed to identify a pair of b quarks clustered in a single jet\cite{CMS:2016jdj}. The algorithm combines information from displaced tracks and the presence of one or two secondary vertices within the Higgs boson jet in a dedicated multivariate algorithm. The decay chains of the two b hadrons are resolved by associating reconstructed secondary vertices with the directions of the two $N$-subjettiness axes.
Tight and loose operating points are chosen for Higgs boson jets that have corresponding false-positive rates for light quark and gluon jets being identified as jets from b quarks of about 0.8 and 8\%, with efficiencies of approximately 35 and 75\%, respectively.
Scale factors, derived from data in events enriched by jets containing muons\cite{CMS:2016jdj}, are applied to the simulation to correct for the differences between data and simulation. See Figure.~\ref{fig:tau21bb} right plot.




\section{Analysis}
\subsection{$X\to VV\to q\bar{q}q\bar{q}$}
This analysis\cite{CMS:2017skt} presents a search for narrow resonances with W or Z bosons decaying hadronically at resonance masses larger than 1.2 TeV.
The results are applicable to models predicting narrow resonances and are compared to several benchmark models. We consider final states produced when a VV boson pair decays into four quarks or qV decays into three quarks, and each boson is reconstructed as a single jet, resulting in events with two reconstructed jets (dijet channel).

\begin{figure}[!hbtp]\centering
  \includegraphics[width=0.28\textwidth]{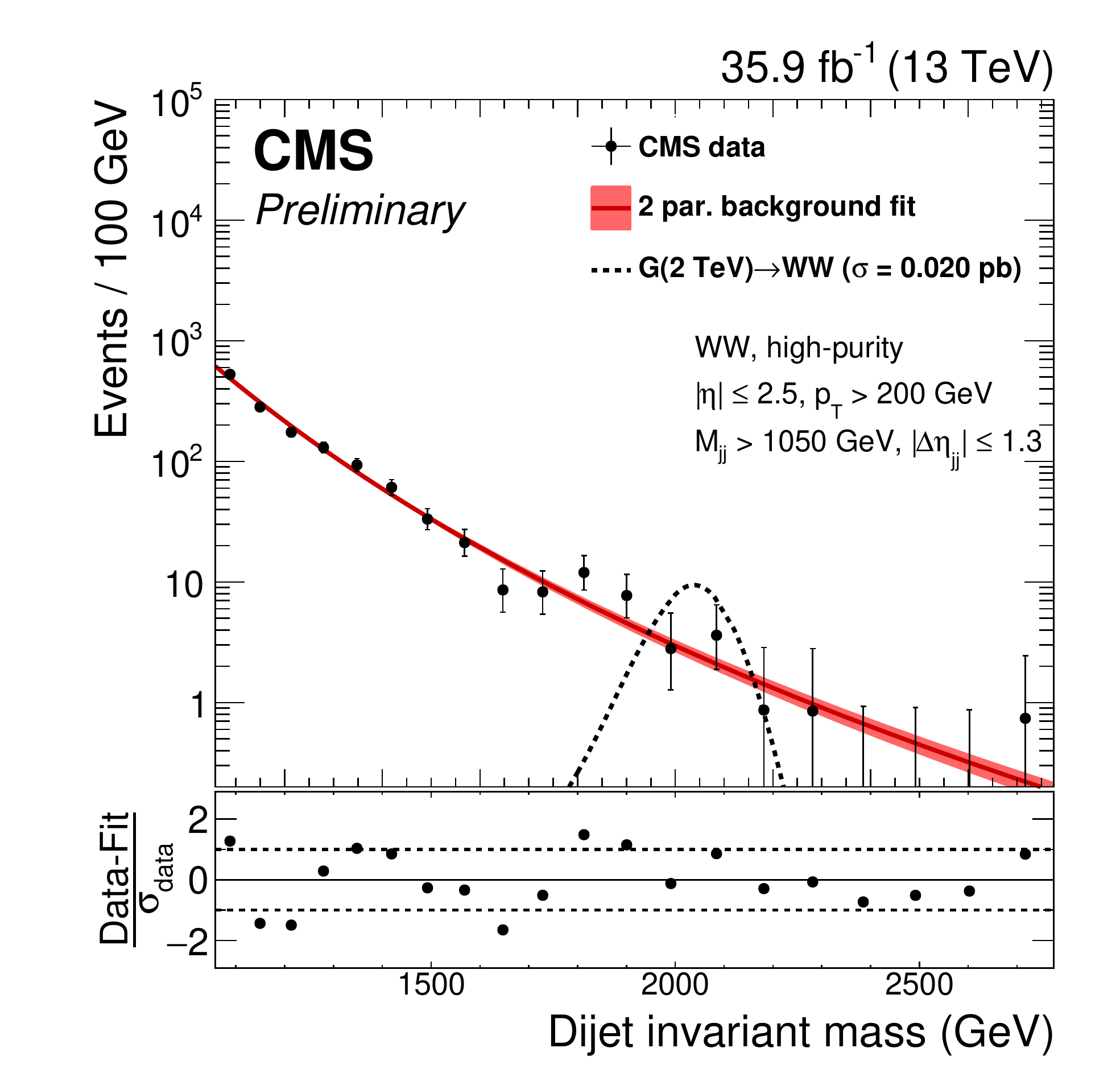}
  \includegraphics[width=0.35\textwidth]{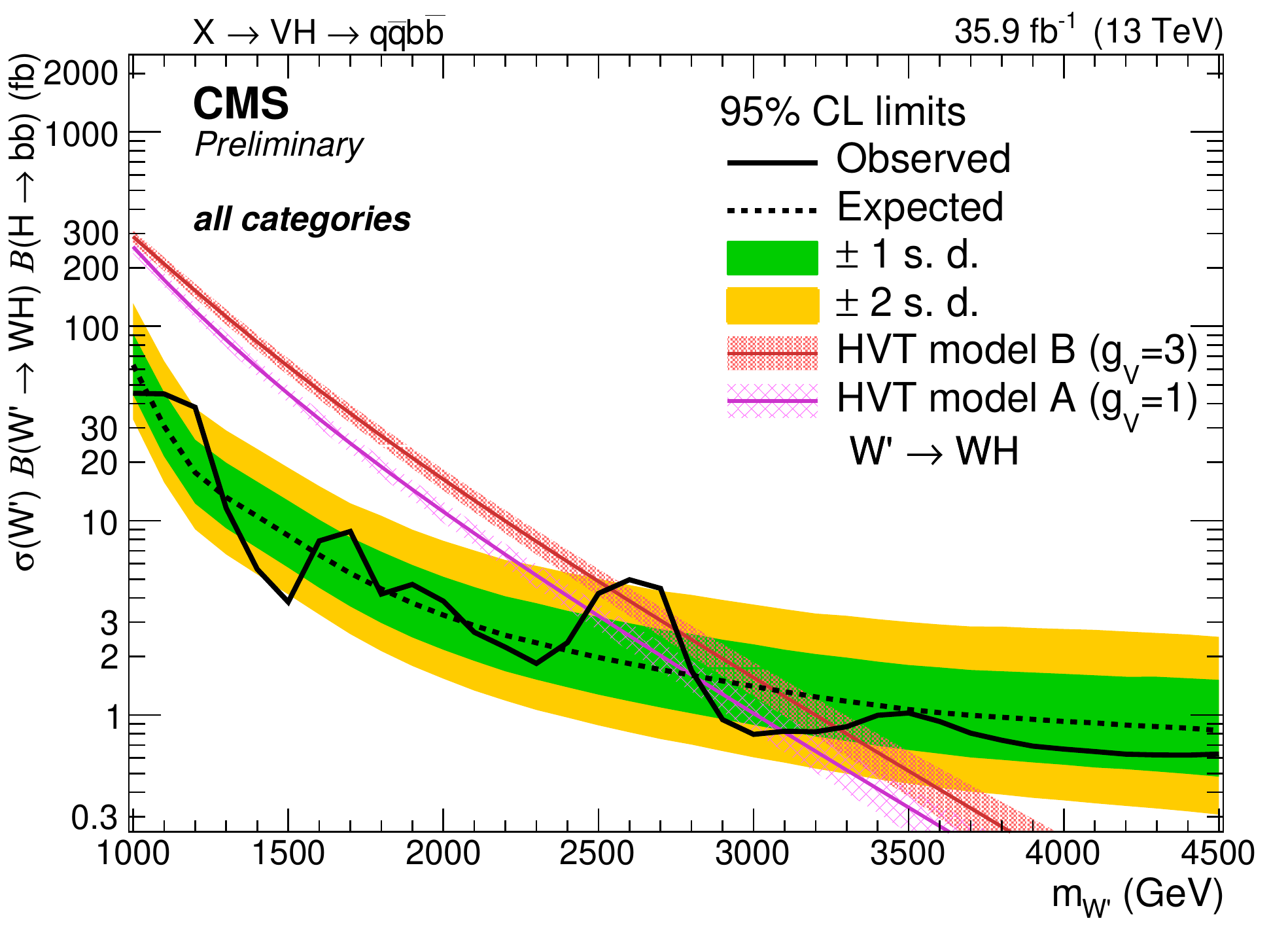}
  \includegraphics[width=0.35\textwidth]{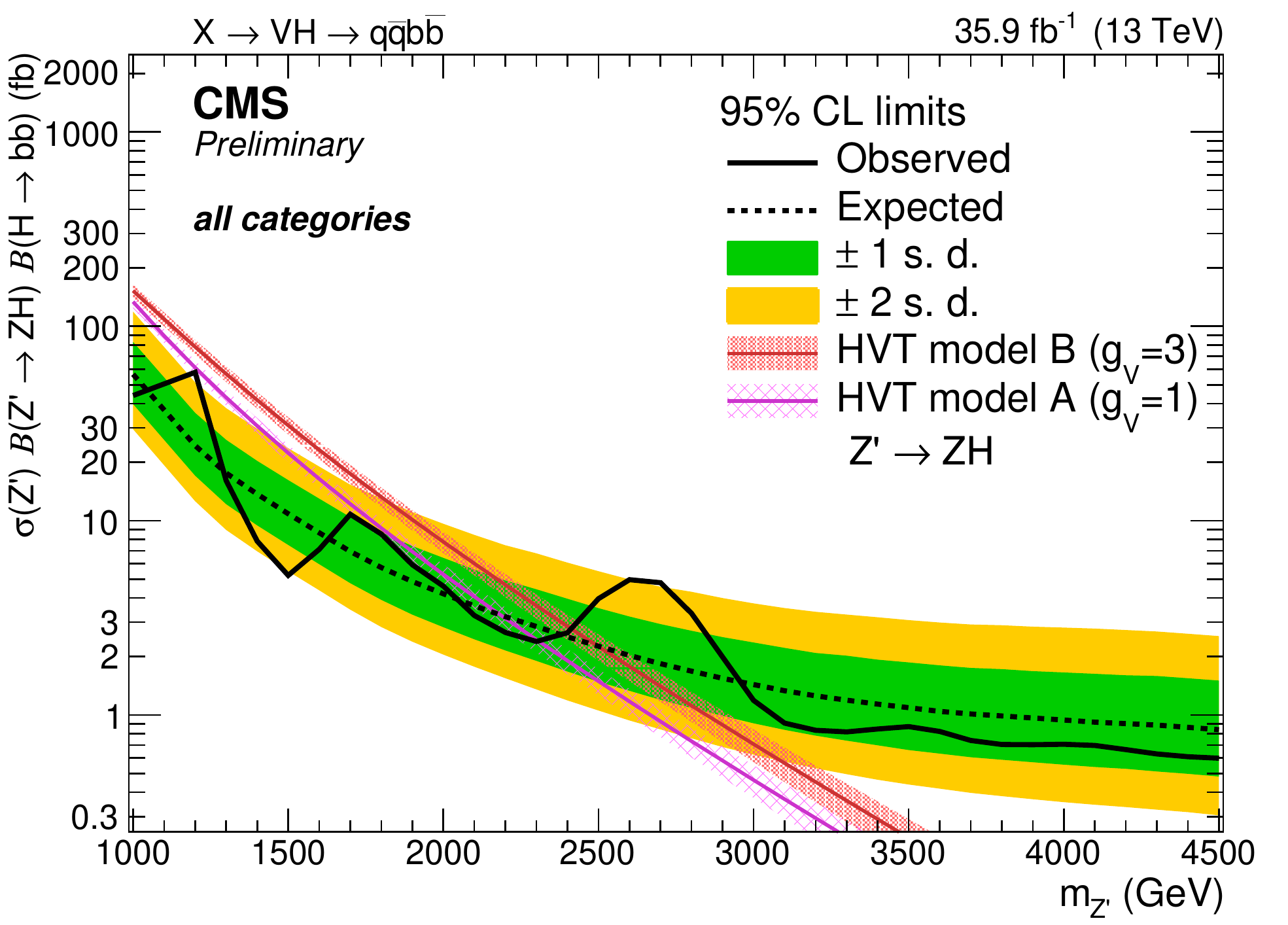}\\
  \caption{
    The dijet invariant mass distribution $m_{jj}$ in data(left) and observed and expected 95\% CL upper limits of $X\to VH\to q\bar{q}b\bar{b}$ on the product $\sigma(X), \mathcal{B}(X\to WH), \mathcal{B}(H\to b\bar{b} )$ (left) and $\sigma(X), \mathcal{B}(X\to ZH), \mathcal{B}(H\to b\bar{b} )$ (right) as a function of the resonance mass for a single narrow spin-1 resonance, for the combination of the eight categories, and including all statistical and systematic uncertainties(right).\cite{CMS:2017skt}
    \label{fig:qqqq} }
\end{figure}

The analysis exploits the large branching fraction of vector boson decays to quark final states. Due to the large masses of the studied resonances, the boson decay products are highly collimated and reconstructed as single, large-radius jets. Jet substructure techniques, referred to as jet \emph{V tagging}  are employed to suppress the SM backgrounds, which largely arise from the hadronization of single quarks and gluons. 
The $m_{jj}$~distributions observed in data are dominated by SM background processes, which in turn are dominated by multijet production where quark or gluon jets are falsely identified as V jets. Additional subdominant backgrounds include W and Z boson production, top quark pair production, single top quark production, and non-resonant diboson processes. Those backgrounds are estimated from simulation to each contribute less than about 3\% of the total number of background events in the signal region and are therefore not separated in the background estimation.

We assume that the multijet SM background can be described by a smooth, monotonically decreasing distribution, which can be parametrized. The search is performed by fitting the sum of the analytical functions for background and signal to the whole dijet spectrum in data. Separate fits are made for each signal mass hypothesis and each analysis category. Neither data control regions nor simulated background samples are used directly by this method. The background functions are of the form:
%
\begin{equation}
\label{eq:dijet1}
\frac{dN}{dm_{jj}}= \frac{ P_0(1-m_{jj}/\sqrt{s})^{P_2} } { (m_{jj}/\sqrt{s})^{P_1} }\:\:\textrm{(3-par. form)},
\quad\quad\quad\quad
\frac{dN}{dm{jj}}= \frac{ P_0 } { (m_{jj}/\sqrt{s})^{P_1} }\:\:\textrm{(2-par. form)},
\end{equation}
where $m_{jj}$ is the dijet invariant mass (equivalent to the diboson or quark-boson candidate mass $m_{VV}$ or $m_{qV}$ for the signal),
$\sqrt{s}$ is the center-of-mass energy,
$P_0$ is a normalization parameter for the probability density function, and $P_1$ and $P_2$ describe the shape.
Starting from the two-parameter functional form, a Fisher F-test \cite{FisherTest} is used to check at 10\% confidence level,
if additional parameters are needed to model the individual background distribution.

No evidence is found for a signal and upper limits on the resonance production cross section are set
as function of the resonance mass. The results are interpreted in the
context of the bulk graviton model, heavy vector triplet $W'$ and $Z'$ resonances, and excited quark resonances ${q^*}$.
For the heavy vector triplet model~B, we exclude $W'$ and $Z'$ resonances with masses below 3.2 and 2.7TeV, respectively.

\subsection{$X\to VH\to q\bar{q}b\bar{b}$}

This analysis\cite{CMS:2017eme} describes the search in proton-proton collisions at 13TeV for heavy resonances decaying to final states containing a SM vector boson and a Higgs boson, which subsequently decay into a pair of quarks and a pair of b quarks, respectively. Use of the hadronic decay modes takes advantage of the large branching fractions, which compensate for the effect of the large multijet background.

This analysis has similar topology and background estimation to VV resonances search, but dedicated identification for $H\to b\bar{b} $ (b-tagging).

Compared to previous measurements, the range of resonance masses excluded within the framework of benchmark model~B of the heavy vector triplet model is extended substantially to values as high as 3.3TeV. More generally, the results lead to a significant reduction in the allowed parameter space for heavy vector triplet models(\ref{fig:qqqq} middle and right plots).

\subsection{$X\to HH\to b\bar{b}b\bar{b}$}

The search\cite{CMS:2017gxe} for heavy narrow resonances decaying to a $HH$ pair is presented in this subsection. Both Higgs bosons are required to decay via $b\bar{b}$. When the decaying resonance mass is sufficiently large, the two Higgs bosons are produced with Lorentz boost considerably higher than their mass, and each $b\bar{b}$ pair is merged into one single jet. Jet algorithms with large distance parameters are used to reconstruct the $H\to b\bar{b}$ decays. The $H$ jet candidates are then identified by exploiting a combination of jet substructure and $b$ tagging techniques. The search is then performed by looking for a resonance in the dijet mass distribution in a mass range of 750--3000GeV. The background is mostly due to SM multijet production and it is entirely estimated from data.

For the purpose of background estimation, the $M_{jj}^{red}$ spectrum is divided into two ranges, between 800 and 1200GeV, and above 1200GeV. The background estimation in both ranges relies on a set of sidebands that are used to predict the total background normalization. 
Above 1200 GeV, the $M_{jj}^{red}$ shape is monotonically falling and hence a smoothness test of the background is used to search for localized excesses, using a parametric functional form to fit the background, while the normalization is constrained from the sidebands.

The different sidebands are defined by two variables related to the leading-$p_{T}$ jet: (i) its soft-drop mass $M_{j1}$ and (ii) the value of the discriminator of the double-b tagger. The background is estimated in bins of the $M_{jj}^{red}$ distribution. Considering these two variables, several regions are defined.
The $pre-tag$ region is defined by all events that fulfill all selection requirements without any requirements on neither $M_{j1}$ nor the double-b tagger discriminator of the leading-$p_{T}$ jet. The $signal$ region is the subset of pre-tag events where $M_{j1}$ is inside the $H$ jet mass window, and with the double-b tagger discriminator greater than 0.3. The $anti-tag$ region consists of events with $M_{j1}$ within the $H$ jet mass window, but  with double-b tagger less than 0.3. These events are dominated by the multijet background, and are kinematically identical to the multijet background events in the signal region which we seek to characterize.

In the absence of correlation between $M_{j1}$ and the double-b tagger discriminator values, one could measure in the $M_{j1}$ sideband the ratio of the number of events passing and failing the double-b tagger selection, $R_{p/f} \equiv N_{\rm pass}/N_{\rm fail}$, ({\it i.e.}, the ``pass-fail ratio"), and  scale the yield in the anti-tag region (in each $M_{jj}^{red}$ bin) by this ratio to obtain an estimate of the background normalization in the signal region. To account for the small correlation between the double-b tagger discriminator and $M_{j1}$, the pass-fail ratio is measured as a function of $M_{j1}$. Thus, extending the ``ABCD'' method with more sidebands defines the ``Alphabet'' method. The measurements of $R_{p/f}$ in the $M_{j1}$ sidebands are fit with a quadratic function, where the coefficients are obtained from the fit, and their correlated uncertainties are propagated as the fit uncertainty. This way the fit interpolates the pass-fail ratio through the signal region in $M_{j1}$, and every event in the anti-tag region is scaled by the appropriate pass-fail ratio given the corresponding $M_{j1}$. The double-b tagger mistag efficiency shows no dependence on the jet $p_{T}$, for this reason the measured pass-fail ratio can be used in the entire range of $M_{jj}^{red}$. The Alphabet background estimation is used for a resonance mass $M(X) < 1200GeV$. 

\begin{figure}[!hbtp]\centering
  \includegraphics[width=0.32\textwidth]{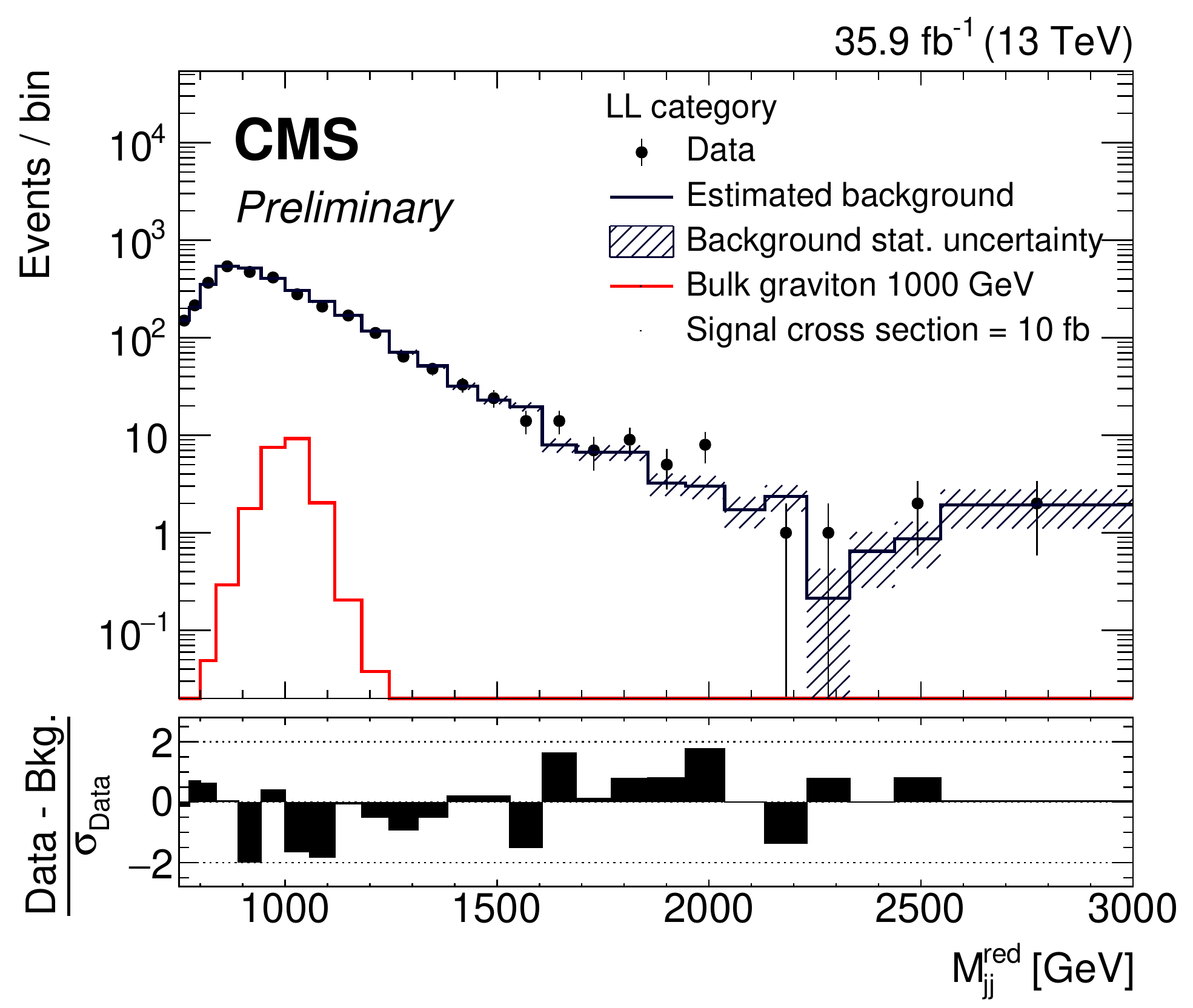}
  \includegraphics[width=0.32\textwidth]{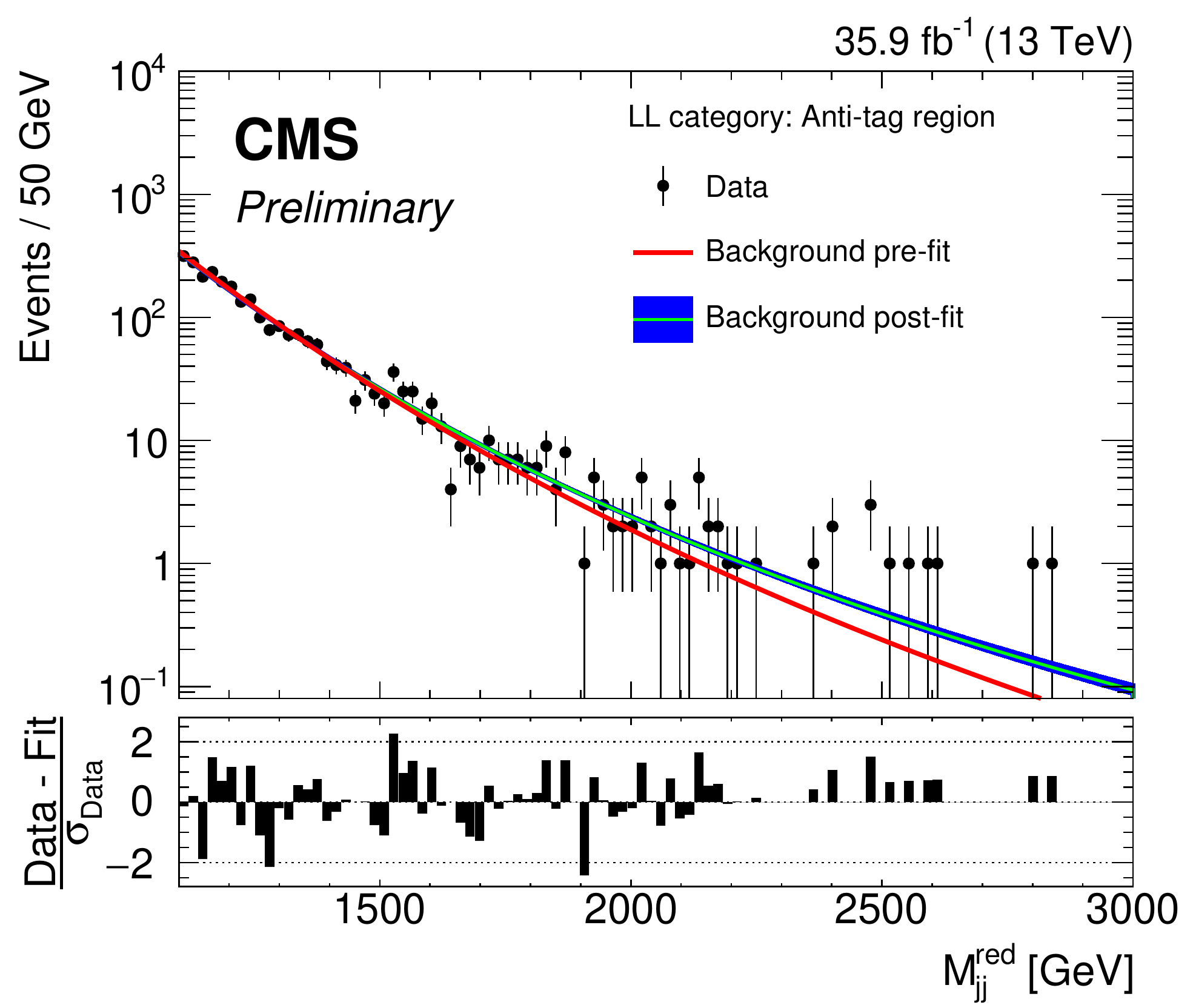}
  \includegraphics[width=0.32\textwidth]{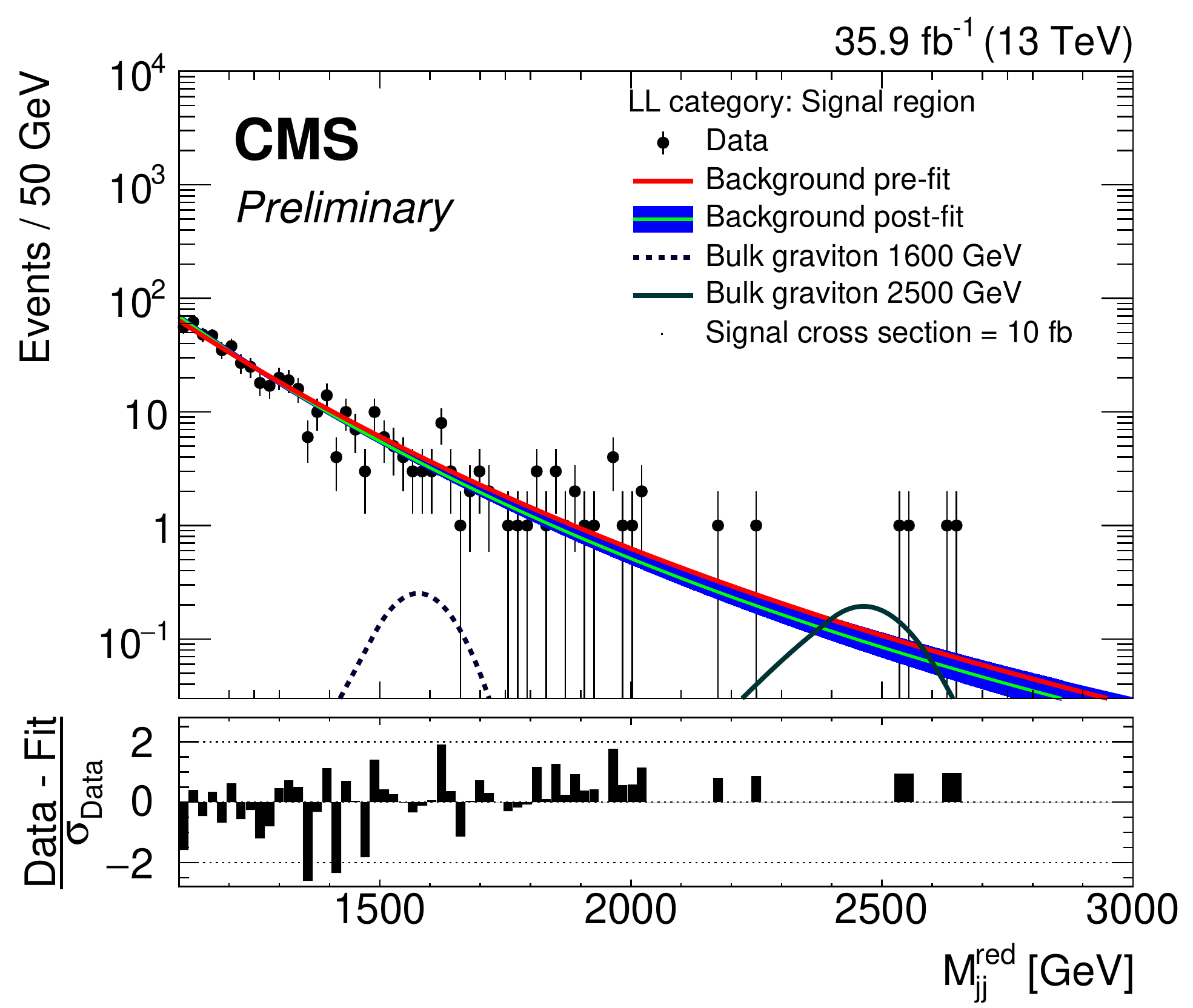}\\
  \caption{
Comparison between the data and the predicted background using the ``Alphabet'' method for the LL (left), the $M_{jj}^{red}$ distributions in the anti-tag region for the LL (middle) and The $M_{jj}^{red}$ distributions in the the signal region for the LL (right).\cite{CMS:2017gxe}
    \label{fig:bbbb} }
\end{figure}

In the absence of such an excess, upper limits are set on the production cross section times the branching fraction of a Kaluza-Klein bulk graviton and a Randall-Sundrum radion decaying to a pair of standard model Higgs bosons, for various hypothetical masses of the bulk graviton and the radion in the range 800--3000GeV. For the mass scale $\Lambda R = 3TeV$, we exclude a radion of mass between 970 and 1450GeV.

\subsection{$X\to ZZ\to l^{+}l^{-}\nu\bar{\nu}$}

This analysis\cite{CMS:2017wsr} presents a search for new resonances decaying to ZZ in which one of the Z bosons decays into two charged leptons and the other into two neutrinos.

The most significant background process is the Z+jets resonant production
in Drell-Yan (DY) processes, where a Z boson or the hadrons recoiling
against it are not reconstructed perfectly, producing a signal-like
final state with $E_{\mathrm{T}}^{\mathrm{miss}}$ arising from primarily instrumental effects.
Other important sources of background include the non-resonant production of
$ll$ final states and $E_{\mathrm{T}}^{\mathrm{miss}}$, primarily comprising $\mathrm{t\bar{t}}$ and WW production
with leptonic final states, and the irreducible background from SM
ZZ/WZ production.
The $E_{\mathrm{T}}^{\mathrm{miss}}$ of the Z+jets background
is instrumental and dominant in the low $E_{\mathrm{T}}^{\mathrm{miss}}$ region,
but can be better separated from signal with the high $E_{\mathrm{T}}^{\mathrm{miss}}$ originating
from the highly boosted invisible Z boson.

The description of real detector effects of Z+jets resonant production
in Drell-Yan (DY) processes is improved by constructing
a data-driven background estimation using $\gamma$-jets data with a reweighting
procedure to reproduce the kinematics of the Z boson in Z+jets events. However, the irreducible background (Reson.) arises mainly the from SM
$\mathrm{q\bar{q} \to ZZ \to } 2l2\nu$ process and also includes a smaller
contribution from WZ and ttZ decays.  This is modeled using MC
samples. Finally, the least non-resonant backgrounds (Non-Reson.), mostly $\mathrm{t\bar{t}}$ and WW, can be significant in regions of
large $E_{\mathrm{T}}^{\mathrm{miss}}$  due to the presence of neutrinos in the final state.
A data driven method is used in the analysis to more precisely model
this background.

As the analysis presented before, still no excess under 1.5TeV.

\begin{figure}[!hbtp]\centering
  \includegraphics[width=0.32\textwidth]{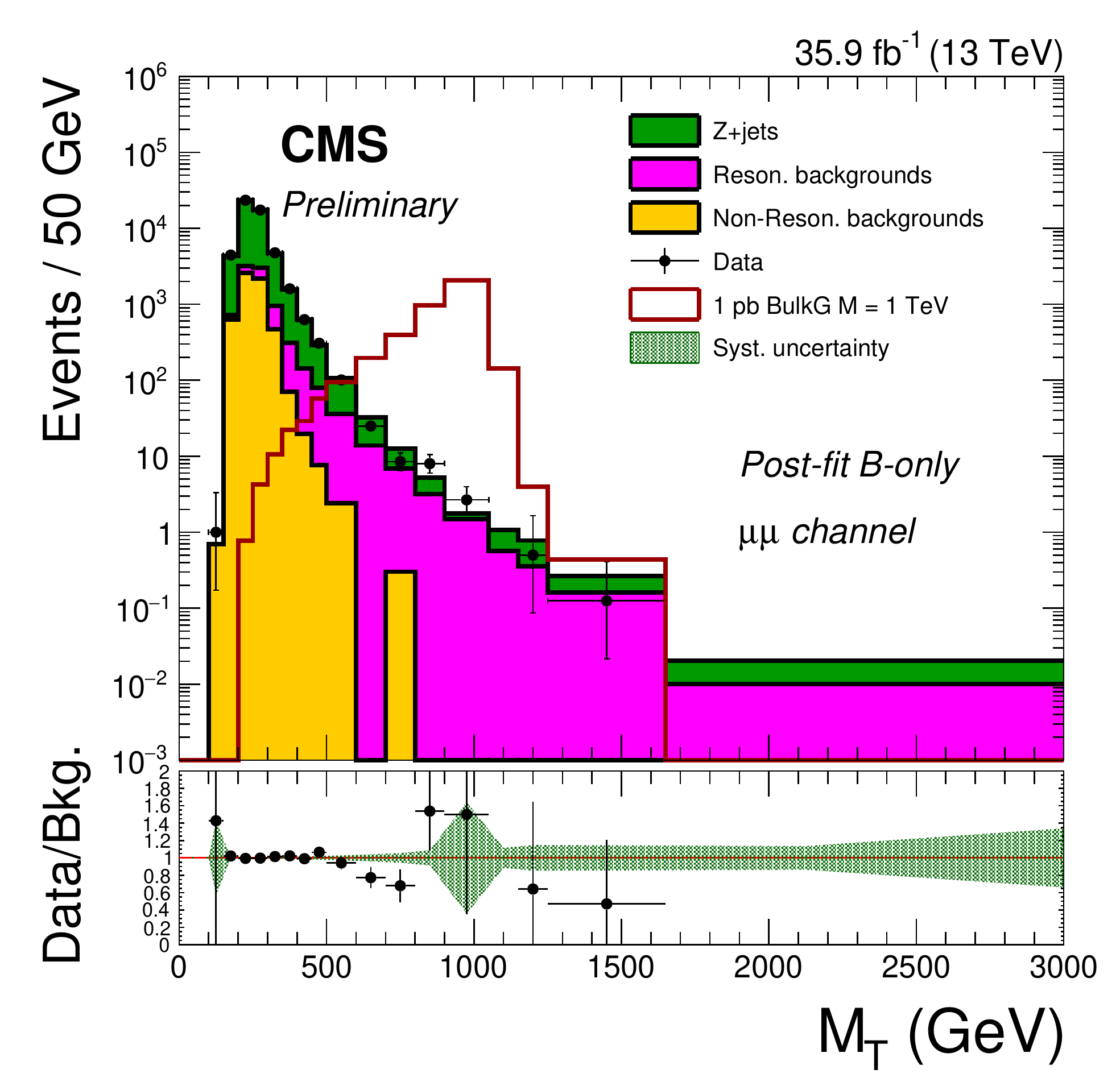}
  \includegraphics[width=0.32\textwidth]{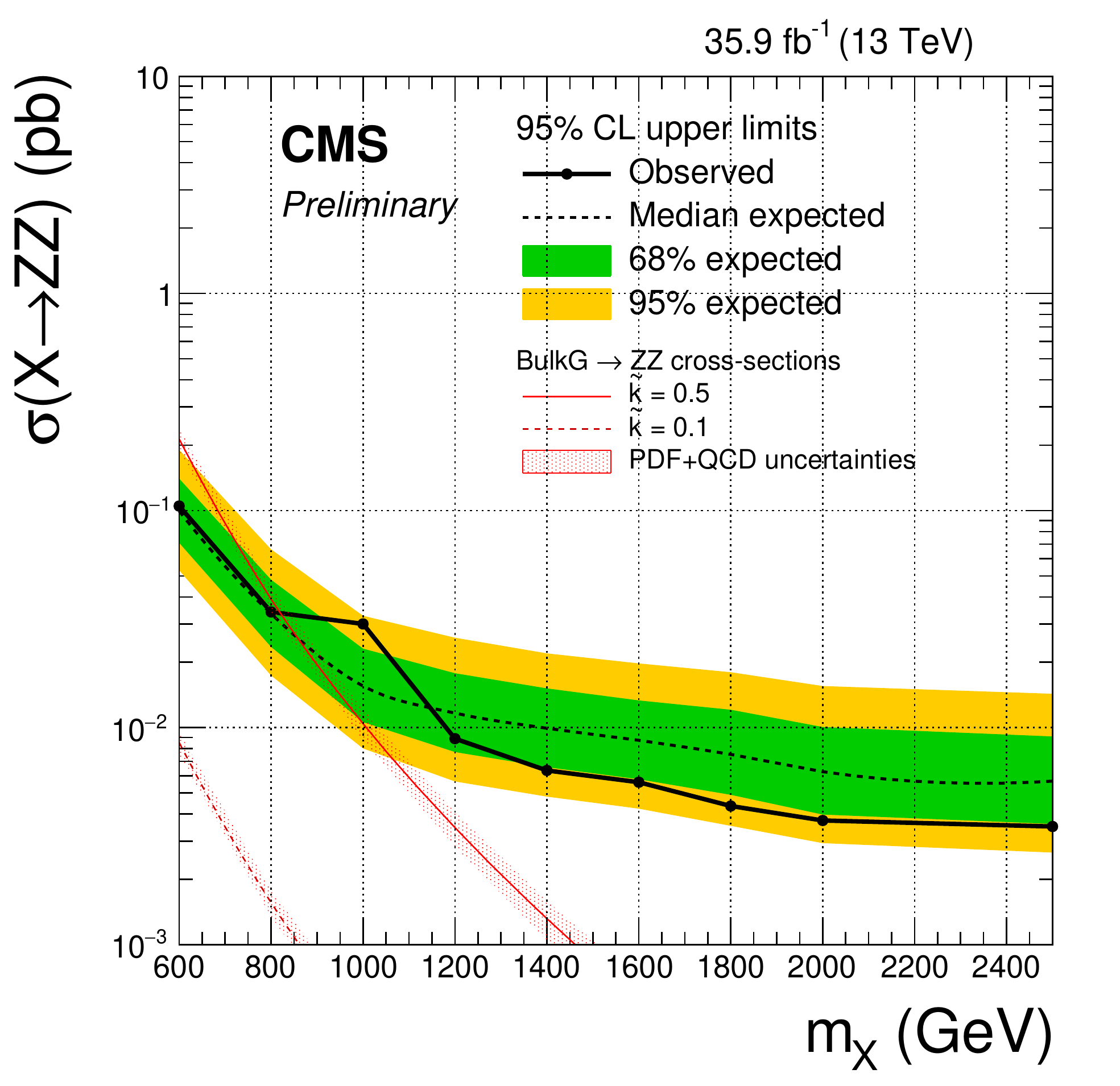}
  \caption{
The $M_{\mathrm{T}}$ distributions for muon ($\mu\mu$) channels
    comparing data and data-driven background modeling after fitting only the background modeling to the data (Post-fit B-only) (left) and Expected and observed limits on the production cross-section of a new spin-2 heavy resonance $\mathrm{X \to ZZ}$ assuming zero mass width separately for the electron ($ee$, left) and muon ($\mu\mu$, right) channels.\cite{CMS:2017wsr}
\label{fig:bbbb} }
\end{figure}

\subsection{$X\to WV\to l\nu q\bar{q}$ and $X\to ZV\to ll q\bar{q}$}

The two analyses\cite{CMS:2016pfl, CMS:2017mrw} use $12.9~\mathrm{fb}^{-1}$ data collected from CMS before ICHEP2016. 
They have similar background estimation method($\alpha-method$) and one AK8 jet decayed from one Vector boson. The differences are that for $X\to WV\to l\nu q\bar{q}$ search, W decay to one lepton and one neutrino, and for $X\to ZV\to ll q\bar{q}$ search, the Z boson decay to two leptons with opposite charge.
Conclusively, no excess is observed.

\subsection{Combination results by 2015 data}

This paper\cite{Sirunyan:2017nrt} combines results from the following final states:
3$l\nu$ (8TeV);
$llq\bar{q}$ (8TeV);
$l\nu q\bar{q}$ (8TeV);
$q\bar{q}q\bar{q}$ (8TeV);
$l\nu b\bar{b}$ (8TeV);
$q\bar{q} \tau\tau $ (8TeV);
$q\bar{q}b\bar{b}$ and $6q$ (8TeV);
$l\nu q\bar{q}$ (13TeV);
$q\bar{q}q\bar{q}$ (13TeV);
and $llb\bar{b}$, $l\nu b\bar{b}$, and $\nu \nu b\bar{b}$ (13TeV).
Since some more forward parts of the detector, which provide information for the calculation of the missing transverse momentum, were not in optimal condition for a fraction of the 2015 data-taking period, the analyses of 13TeV data in the $l\nu q\bar{q}$, $l\nu b\bar{b}$ and $\nu\nu b\bar{b}$ decay channels are based on the dataset corresponding to the integrated luminosity of $2.3~\mathrm{fb}^{-1}$ rather than $2.7~\mathrm{fb}^{-1}$.
In fact, 2016 searches already have more sensitive than combination results.
\section{Conclusions}

Searching for heavy resonances is one of the most direct ways to find new
physics at TeV scale. Significant developments in boosted object techniques have also taken place. From the list of the analyses, we could see a rich phenomenology and  final states from VV, VH, HH decays which have clear experimental signatures and allows crosschecks among different channels. However, no significant excess has been observed in recent 13 TeV results shown in this paper. We look forward to more diboson results with 2016-17 LHC data.

\Acknowledgements
I would like to thank the LHCP2017 organizers for their hospitality and
the wonderful working environment. I acknowledge the support from
the National Natural Science Foundation of China, under Grants No.11661141008.


\begin{thebibliography}{99}


  
  

\bibitem{Fitzpatrick:2007qr}
  A.~L.~Fitzpatrick, J.~Kaplan, L.~Randall and L.~T.~Wang,
  JHEP {\bf 0709} (2007) 013
  doi:10.1088/1126-6708/2007/09/013
  [hep-ph/0701150].


\bibitem{Randall:1999ee}
  L.~Randall and R.~Sundrum,
  Phys.\ Rev.\ Lett.\  {\bf 83} (1999) 3370
  doi:10.1103/PhysRevLett.83.3370
  [hep-ph/9905221].

\bibitem{Pappadopulo:2014qza}
  D.~Pappadopulo, A.~Thamm, R.~Torre and A.~Wulzer,
  JHEP {\bf 1409} (2014) 060
  doi:10.1007/JHEP09(2014)060
  [arXiv:1402.4431 [hep-ph]].

\bibitem{Chatrchyan:2008aa}
  S.~Chatrchyan {\it et al.} [CMS Collaboration],
  JINST {\bf 3} (2008) S08004.
  doi:10.1088/1748-0221/3/08/S08004


\bibitem{CMS:2014ata}
  CMS Collaboration,
  CMS-PAS-JME-14-001, https://cds.cern.ch/record/1751454.

\bibitem{Bertolini:2014bba}
  D.~Bertolini, P.~Harris, M.~Low and N.~Tran,
  JHEP {\bf 1410} (2014) 059
  doi:10.1007/JHEP10(2014)059
  [arXiv:1407.6013 [hep-ph]].

\bibitem{Larkoski:2014wba}
  A.~J.~Larkoski, S.~Marzani, G.~Soyez and J.~Thaler,
  JHEP {\bf 1405} (2014) 146
  doi:10.1007/JHEP05(2014)146
  [arXiv:1402.2657 [hep-ph]].

\bibitem{Thaler:2010tr}
  J.~Thaler and K.~Van Tilburg,
  JHEP {\bf 1103} (2011) 015
  doi:10.1007/JHEP03(2011)015
  [arXiv:1011.2268 [hep-ph]].

\bibitem{CMS:2016jdj}
  CMS Collaboration,
  CMS-PAS-BTV-15-002, https://cds.cern.ch/record/2195743.

\bibitem{CMS:2017skt}
  CMS Collaboration,
  CMS-PAS-B2G-17-001, https://cds.cern.ch/record/2256663.

\bibitem{FisherTest}
R. G. Lomax and D. L. Hahs-Vaughn,
"Statistical concepts: a second course",
Routledge Academic,
London,
2007

\bibitem{CMS:2017eme}
  CMS Collaboration,
  CMS-PAS-B2G-17-002, https://cds.cern.ch/record/2256742.

\bibitem{CMS:2017gxe}
  CMS Collaboration,
  CMS-PAS-B2G-16-026, https://cds.cern.ch/record/2264684.

\bibitem{CMS:2017wsr}
  CMS Collaboration,
  CMS-PAS-B2G-16-023, https://cds.cern.ch/record/2264700.

\bibitem{CMS:2016pfl}
  CMS Collaboration,
  CMS-PAS-B2G-16-020, http://cds.cern.ch/record/2205880.

\bibitem{CMS:2017mrw}
  CMS Collaboration,
  CMS-PAS-B2G-16-022, https://cds.cern.ch/record/2242955.

\bibitem{Sirunyan:2017nrt}
  A.~M.~Sirunyan {\it et al.} [CMS Collaboration],
  arXiv:1705.09171 [hep-ex].


\end{thebibliography}
\end{document}